\RequirePackage{ifpdf}
\ifpdf % We are running pdfTeX in pdf mode
\documentclass[pdftex]{sigma}
\else
\documentclass{sigma}
\fi

\def\EE{\mathcal{E}}
\def\TT{\mathcal{T}}

\begin{document}

\allowdisplaybreaks

\renewcommand{\thefootnote}{$\star$}

\renewcommand{\PaperNumber}{034}

\FirstPageHeading

\ShortArticleName{The Lax Integrable Dif\/ferential-Dif\/ference
Dynamical Systems on Extended Phase Spaces}

\ArticleName{The Lax Integrable Dif\/ferential-Dif\/ference\\ Dynamical
Systems on Extended Phase Spaces\footnote{This
paper is a contribution to the Proceedings of the Eighth
International Conference ``Symmetry in Nonlinear Mathematical
Physics'' (June 21--27, 2009, Kyiv, Ukraine). The full collection
is available at
\href{http://www.emis.de/journals/SIGMA/symmetry2009.html}{http://www.emis.de/journals/SIGMA/symmetry2009.html}}}

\Author{Oksana Ye. HENTOSH}

\AuthorNameForHeading{O.Ye. Hentosh}

\Address{Institute for Applied Problems of Mechanics and Mathematics, \\
National Academy of Sciences of Ukraine,
3B Naukova Str., Lviv, 79060, Ukraine}
\Email{\href{mailto:ohen@ua.fm}{ohen@ua.fm}}

\ArticleDates{Received November 16, 2009, in f\/inal form February 24, 2010;  Published online April 17, 2010}

\Abstract{The Hamiltonian representation for the hierarchy of
Lax-type f\/lows on a dual space to the Lie algebra of shift
operators coupled with suitable eigenfunctions and adjoint
eigenfunctions evolutions of associated spectral problems is found
by means of a specially constructed B\"acklund transformation. The
Hamiltonian description for the corresponding set of squared eigenfunction
symmetry hierarchies is represented. The relation of these
hierar\-chies with Lax integrable $(2+1)$-di\-men\-sion\-al
dif\/ferential-dif\/ference systems and their triple Lax-type
linearizations is analysed. The existence problem of a Hamiltonian
representation for the coupled Lax-type hierarchy on a dual space
to the central extension of the shift operator Lie algebra is
solved also.}


\Classification{37J05; 37K10; 37K30; 37K35; 37K60}

\section{Introduction}
The f\/irst papers on the Lie-algebraic interpretation of Lax
integrable dif\/ferential-dif\/ference systems were published
by~Adler~\cite{hentosh1}, Kostant~\cite{shentosh11} and
Symes~\cite{shentosh19}. They considered non-periodic lattices of
a Toda type~\cite{shentosh7,shentosh16,shentosh17,shentosh18}
related to coadjoint orbits of solvable matrix Lie algebras.

The $\mathcal{R}$-matrix
approach~\cite{hentosh_Bl,bhentosh6,hentosh9,hentosh_Myk,shentosh17,hentosh8}
being useful for the Lie-algebraic description of Lax integrable
nonlinear dynamical systems on functional
manifolds~\cite{hentosh16,hentosh2} turned out to be suitable for
the Lie-algebraic description of Lax integrable
$(1+1)$-dimensional lattice and nonlocal dif\/ferential-dif\/ference
systems by means of the Lie algebra of shift
operators~\cite{shentosh1,hentosh_Bog,shentosh_Kup,shentosh14,hentosh_Oevel1}.

The Lax integrable $(2+1)$-dimensional dif\/ferential-dif\/ference
systems were obtained via the Sato procedure~\cite{hentosh20}
in~\cite{shentosh15,shentosh20,hentosh_Zeng1} whereas in
papers~\cite{shentosh2,shentosh3,shentosh5,bhentosh10,shentosh10,hentosh4}
such dif\/ferential-dif\/ference systems were considered as
Hamiltonian f\/lows on the dual spaces to the central extensions by
the Maurer--Cartan 2-cocycle of shift
operator Lie algebras.

Taking into account that every f\/low from the Lax-type hierarchy on
the dual space to the shift operator Lie algebra or its central
extension can be written as a compatibility condition of the
spectral relationship for the corresponding operator and the
suitable eigenfunction evolution an important problem of f\/inding
the Hamiltonian representation for the hierarchy of Lax-type
f\/lows coupled with the evolutions of eigenfunctions and appropriate
adjoint eigenfunctions naturally arises. In the case when the
spectral relationship admits a f\/inite set of eigenvalues it was
partly solved in the
papers~\cite{bhentosh8,bhentosh27,shentosh9,hentosh10,hentosh12}
for the Lie algebra of integral-dif\/ferential
operators~\cite{hentosh10,hentosh12} and its
supergeneralizations~\cite{bhentosh8,shentosh9} as well as for the
corresponding central extension~\cite{bhentosh27} by means of the
variational property of Casimir functionals under some B\"acklund
transformation.

Section~\ref{section2} deals with a general Lie-algebraic scheme for
constructing the hierarchy of Lax-type f\/lows as Hamiltonian ones
on a dual space to the Lie algebra of shift
operators~\cite{shentosh1,shentosh5}.

In Section~\ref{section3} the Hamiltonian structure for the related coupled
Lax-type hierarchy is obtained by means of the B\"acklund
transformation technique developed
in~\cite{bhentosh8,bhentosh27,shentosh9,hentosh12}.

In Section~\ref{section4} the corresponding hierarchies of squared
eigenfunction
symmetries~\cite{hentosh_ANP,hentosh_Carillo1,hentosh_Carillo2,hentosh_NP,hentosh_Zeng1,hentosh_Zeng2}
for the coupled Lax-type f\/lows are established to be Hamiltonian
also. It is proved that the additional hierarchy of Hamiltonian
f\/lows is generated by the Poisson structure being obtained from
the tensor product of the $\mathcal{R}$-deformed canonical
Lie--Poisson bracket with the standard Poisson bracket on the
related eigenfunctions and adjoint eigenfunctions
superspace~\cite{bhentosh8,bhentosh27,shentosh9,hentosh12} and the
corresponding natural powers of a suitable eigenvalue are their
Hamiltonians. These hierarchies are applied to constructing Lax
integrable $(2+1)$-di\-men\-sion\-al dif\/ferential-dif\/ference
systems and their triple Lax-type linearizations.

In Section~\ref{section5} the results obtained in Section~\ref{section3} are generalized for
the centrally extended Lie algebra of shift
operators~\cite{shentosh2,shentosh5}.

\section[The Lie-algebraic structure of Lax integrable
(1+1)-dimensional differential-difference systems]{The Lie-algebraic structure of Lax integrable\\
$\boldsymbol{(1+1)}$-dimensional dif\/ferential-dif\/ference systems}\label{section2}

Let us consider the Lie algebra $\mathcal{G}$ of linear
operators~\cite{shentosh1,hentosh_Bog,shentosh_Kup,shentosh14,hentosh_Oevel1}
%%%1
\begin{gather}\label{ohen1a}
A: = \EE^m + \sum_{j< m,\ j\in \mathbb{Z}} a_j(n)\EE ^j,\qquad m\in
\mathbb{N},
\end{gather}
where coef\/f\/icients $a_j$ belong to the Schwarz space $\mathcal{S}
(\mathbb{Z};\mathbb{C})$ of quickly decreasing sequences, $j\in
\mathbb{Z}$, which are generated by the shift operator~$\EE$,
satisfying the following rule
\[
\EE ^j a=(\EE ^j a)\EE ^j,
\]
with the standard commutator
\[
[A,B]=AB-BA, \qquad A,B\in
\mathcal{G}.
\]
On the Lie algebra $\mathcal{G}$ there exists the ad-invariant
nondegenerate symmetric bilinear form:
\begin{gather}\label{ohen1}
(A,B):=  \mbox{Tr}\, (AB), \qquad A,B\in \mathcal{G},
\end{gather}
where
\[
\mbox{Tr}\,A:=\sum_{n\in \mathbb{Z}}a_0 (n)
\]
for any operator $A\in \mathcal{G}$ in the form~\eqref{ohen1a}.

With taking into account~\eqref{ohen1} the dual space to the Lie
algebra $\mathcal{G}$ can be identif\/ied with the Lie algebra, that
is, $\mathcal{G}^*\simeq \mathcal{G}$. The linear subspaces
$\mathcal{G}_+\subset \mathcal{G}$ and $\mathcal{G}_- \subset
\mathcal{G}$ such as
%%%3
\begin{gather}
\mathcal{G}_+ := \left\{ a= \EE^m + \sum_{0\le j < m,\ j\in
\mathbb{Z}} a_j \EE ^j:\ \ a_j\in \mathcal{S}
(\mathbb{Z};\mathbb{C}),\ m\in \mathbb{N} \right\} ,
\nonumber \\
\mathcal{G}_-:= \left\{ b=\sum_{\ell\in \mathbb{N}}^{\infty} b_j
\EE ^{-\ell} :\ \ b_j\in \mathcal{S}
(\mathbb{Z};\mathbb{C})\right\} , \label{ohen2}
\end{gather}
are Lie subalgebras in $\mathcal{G}$ and
\[
\mathcal{G} =\mathcal{G}_+\oplus \mathcal{G}_-.
\]
The following spaces can be identif\/ied as
\[
\mathcal{G}_+ ^*\simeq \mathcal{G}_- \circ \EE, \qquad
\mathcal{G}_- ^*\simeq \mathcal{G}_+ \circ \EE .
\]
Owing to the splitting $\mathcal{G}$ into the direct sum of its Lie
subalgebras \eqref{ohen2}, one can construct a~Lie--Poisson
structure on $\mathcal{G}^*$ by use of the special linear
endomorphism $\mathcal{R}$ of $\mathcal{G}$:
\[
\mathcal{R} =(P_+-P_-)/2, \qquad P_{\pm}\mathcal{G}
=\mathcal{G}_{\pm}, \qquad \ P_{\pm}\mathcal{G}_{\mp}=0.
\]
The Lie--Poisson bracket on $\mathcal{G}^*$ is given as
\begin{gather}\label{ohen3}
\left\{ \gamma,\mu \right\}_{\mathcal{R}}(l)= \left (
l,[\nabla\gamma(l),\nabla\mu(l)]_{\mathcal{R} }  \right ),
\end{gather}
where $\gamma,\mu\in \mathcal{D}(\mathcal{G}^*)$,
$\mathcal{D}(\mathcal{G}^*)$ is a space of Frechet-smooth
functionals on $\mathcal{G}^*$, $l\in\mathcal{G}^*$ and for all
$A,B\in \mathcal{G}$ the $\mathcal{R}$-deformed commutator has the
form~\cite{hentosh_Bl,bhentosh6,hentosh9,hentosh_Oevel1,hentosh_Myk,shentosh17,hentosh8}
\[
[a, b]_{\mathcal{R}}=[\mathcal{R} a, b]+[a,\mathcal{R} b].
\]
Based on the scalar product~\eqref{ohen1} the gradient
$\nabla\gamma (l)\in \mathcal{G}$ of some functional
$\gamma\in\mathcal{D} (\mathcal{G}^*)$ at the point $l\in
\mathcal{G}^*$ is naturally def\/ined as
\[
\delta\gamma (l)= \left (\nabla \gamma (l),\delta l \right ).
\]

Let $I (\mathcal{G}^*)$ be a set of Casimir functionals on
$\mathcal{G}^*$, being invariant with respect to ${\rm Ad}^*$-action of
the abstract Lie group $G$ corresponding to the Lie algebra
$\mathcal{G}$ by def\/inition. Every Casimir functional $\gamma\in I
(\mathcal{G}^*)$ obeys the following condition at the point $l\in
\mathcal{G}^*$:
%%%5
\begin{gather}\label{ohen5}
[l , \nabla\gamma (l)]=0.
\end{gather}
The relationship~\eqref{ohen5} is satisf\/ied by the hierarchy of
functionals $\gamma_n\in I(\mathcal{G}^*)$, $n\in\mathbb N$,
taking the forms~\cite{shentosh1,hentosh_Bl}
%%%6
\begin{gather}\label{ohen6}
\gamma_n (l)= \frac 1{n+1} (l,l^n).
\end{gather}
The Lie--Poisson bracket \eqref{ohen3} generates the hierarchy of
Hamiltonian dynamical systems on $\mathcal{G}^*$ with Casimir
functionals $\gamma_n\in I(\mathcal{G}^*)$, $n\in \mathbb{N}$, as
Hamiltonian functions:
\begin{gather}\label{ohen9}
d l/dt_n= [\mathcal{R}\nabla\gamma_n (l),  l]= [(\nabla\gamma_n
(l))_+, l] .
\end{gather}
where the subscript ``$+$'' denotes a projection on the Lie
subalgebra~$\mathcal{G}_+$.

The latter equation is equivalent to the usual commutator Lax-type
representation, which can be considered as a compatibility
condition for the spectral problem
\begin{gather}\label{ohen10}
(l  f)= \lambda f,
\end{gather}
where $f\in W:=L_2 (\mathbb{Z};\mathbb{C})$,
$\lambda\in\mathbb{C}$ is a spectral parameter, and the following
evolution equation:
\begin{gather}\label{ohen11}
df/dt_n= ((\nabla\gamma_n (l))_+f).
\end{gather}
The corresponding evolution for the adjoint eigenfunction $f^*\in
W^*\simeq W$ takes the form
%%%10
\begin{gather}\label{ohen12}
df^*/dt_n=-((\nabla\gamma_n (l))^*_+ f^*).
\end{gather}

Further one will assume that the spectral
relationship~\eqref{ohen10} admits $N \in \mathbb{N}$ dif\/ferent
eigenvalues $\lambda_i\in \mathbb{C}$,
$i=\overline{1,N}:=1,\ldots,N$, and study algebraic properties of
equation~\eqref{ohen9} combined with $N\in \mathbb{N}$ copies
of~\eqref{ohen11}:
%%%11
\begin{gather}\label{ohen13}
df_i/dt_n= ((\nabla\gamma_n (l))_+f_i),
\end{gather}
for the corresponding eigenfunctions $f_i\in W$,
$i=\overline{1,N}$, and the same number of copies
of~\eqref{ohen12}:
%%%12
\begin{gather}\label{ohen14}
df_i^*/dt_n=-((\nabla\gamma_n (l))^*_+ f_i^*),
\end{gather}
for the suitable adjoint eigenfunctions $f_i^*\in W^*$, being
considered as a coupled evolution system on the space
$\mathcal{G}^*\oplus W^{2N}$.

\section[The Poisson bracket on the extended phase space of a Lax integrable
(1+1)-dimensional differential-difference system]{The Poisson bracket on the extended phase space of a Lax\\ integrable
$\boldsymbol{(1+1)}$-dimensional dif\/ferential-dif\/ference system}\label{section3}

To give the description below in a compact form one will use the
following notation of the gradient vector:
\[
\nabla \overline\gamma (\tilde l, \tilde {\rm f}, \tilde {\rm
f}^*):= (\delta\overline\gamma/\delta\tilde l,\,
\delta\overline\gamma/\delta\tilde {\rm f},\,
\delta\overline\gamma/\delta\tilde {\rm f}^*)^\top ,
\]
where $\tilde {\rm f}:=(\tilde f_1, \ldots , \tilde f_N)^\top$,
$\tilde {\rm f}^*:=(\tilde f^*_1, \ldots , \tilde f^*_N)^\top\ $
and $\ \delta\overline\gamma/\delta \tilde {\rm
f}:=(\delta\overline\gamma/\delta \tilde f_1, \ldots ,$ $
\delta\overline\gamma/\delta \tilde f_N)^\top$,
$\delta\overline\gamma/\delta \tilde {\rm
f}^*:=(\delta\overline\gamma/\delta f_1^*, \ldots ,
\delta\overline\gamma/\delta \tilde f_N^*)^\top$, at the point
$(\tilde l, \tilde {\rm f}, {\rm f}^*)^\top \in
\mathcal{G}^*\oplus W^{2N}$ for any Frechet-smooth functional
$\overline\gamma\in\mathcal{D} (\mathcal{G}^*\oplus W^{2N})$.

On the spaces $\mathcal{G}^*$ and $W^{2N}$ there exist canonical
Poisson structures in the forms
%%%13
\begin{gather}\label{ohen15}
\delta\overline\gamma/\delta\tilde l
\stackrel{\tilde\theta}{\mapsto} [\tilde
l,(\delta\overline\gamma/\delta\tilde l)_+] - [\tilde
l,\delta\overline\gamma/\delta\tilde l]_{>0} ,
\end{gather}
where $\tilde\theta: \TT ^*(\mathcal{G}^*)\to \TT (\mathcal{G}^*)$
is an implectic operator corresponding to \eqref{ohen3} at the
point $\tilde l\in\mathcal{G}^*$, $A_{>0}:=A_+-A_0$, $A_0:=a_0$,
for an arbitrary operator $A\in \mathcal{G}$ in the
form~\eqref{ohen1a}, and
%%%14
\begin{gather}\label{ohen16}
(\delta\overline\gamma/\delta\tilde {\rm f},\,
\delta\overline\gamma/\delta\tilde {\rm f}^*)^\top
\stackrel{\tilde J}{\mapsto} (-\delta\overline\gamma/\delta\tilde
{\rm f}^*,\, \delta\overline\gamma/\delta\tilde {\rm f})^\top  ,
\end{gather}
where $\tilde J:\TT^*(W^{2N})\to \TT(W^{2N})$ is an implectic
operator corresponding to the symplectic form
$\omega^{(2)}=\sum_{i=1}^N d\tilde f_i^*\wedge d\tilde f_i$ at the
point $(\tilde {\rm f}, \tilde {\rm f}^*)\in W^{2N}$. It should be
noted here that Poisson structure~\eqref{ohen15} generates
equation~\eqref{ohen9} for any Casimir functional $\gamma\in
I(\mathcal{G}^*)$.

Thus, one can obtain a Poisson structure on the extended phase
space $\mathcal{G}^*\oplus W^{2N}$ as the tensor product
$\tilde\Theta :=\tilde\theta \otimes \tilde J$ of \eqref{ohen15}
and \eqref{ohen16}.

To f\/ind a Hamiltonian representation for the coupled dynamical
systems~\eqref{ohen9}, \eqref{ohen13} and \eqref{ohen14} one will
make use of an approach, described in
papers~\cite{bhentosh8,bhentosh27,shentosh9,hentosh_Myk,hentosh12},
and will consider the following B\"acklund transformation:
%%%15
\begin{gather}\label{ohen17}
(\tilde l, \tilde {\rm f}, \tilde {\rm f}^*)^\top
\stackrel{B}{\mapsto} (l(\tilde l, \tilde {\rm f}, \tilde {\rm
f}^*), \, {\rm f} =\tilde {\rm f}, \, {\rm f}^*=\tilde {\rm
f}^*)^\top ,
\end{gather}
generating some Poisson structure $\Theta:
\TT^*(\mathcal{G}^*\oplus W^{2N})\to \TT(\mathcal{G}^*\oplus
W^{2N})$ on $\mathcal{G}^*\oplus W^{2N}$. The main condition
imposed on mapping \eqref{ohen17} is the coincidence of the
resulting dynamical system
%%%16
\begin{gather}\label{ohen18}
(d l/dt_n,\,d {\rm f} /dt_n,\, d {\rm f}^*/dt_n)^\top:= -\Theta \,
\nabla \overline{\gamma}_n ( l, {\rm f} , {\rm f}^*)
\end{gather}
with equations~\eqref{ohen9}, \eqref{ohen13} and \eqref{ohen14} in
the case of $\overline{\gamma}_n \in I (\mathcal{G}^*) $, $n\in
\mathbb{N}$, independent of variables $({\rm f},{\rm f} ^*)\in
W^{2N}$.

To satisfy that condition one will f\/ind a variation of a Casimir
functional $\overline{\gamma}_n:=\left . \gamma_n \right
|_{l=l(\tilde l, {\rm f}, {\rm f}^*)} \in \mathcal{D}
(\mathcal{G}^*\oplus W^{2N}) $, $n\in \mathbb{N}$, under
$\delta\tilde l=0$, taking into account evolutions~\eqref{ohen13},
\eqref{ohen14} and B\"acklund transformation
def\/inition~\eqref{ohen17}. There follows
%%%17
\begin{gather}
\delta\overline{\gamma}_n (\tilde l, \tilde {\rm f},
\tilde {\rm f}^*)\big|_{\delta\tilde l=0} =\sum_{i=1}^N \big(
\langle \delta\overline{\gamma}_n /\delta\tilde f_i,\delta\tilde f_i\rangle +
\langle \delta\overline{\gamma}_n /\delta\tilde f_i^*,\delta\tilde
f_i^*\rangle \big)
\nonumber \\
\phantom{\delta\overline{\gamma}_n (\tilde l, \tilde {\rm f},
\tilde {\rm f}^*)\big|_{\delta\tilde l=0}}{}
= \sum_{i=1}^N   \big( \langle -d\tilde f_i^*
/dt_n,\delta\tilde f_i \rangle  + \langle d\tilde f_i /dt_n,\delta\tilde f_i^*\rangle
\big) \big|_{\tilde {\rm f}={\rm f},\,
\tilde {\rm f}^*={\rm f}^*} \nonumber \\
\phantom{\delta\overline{\gamma}_n (\tilde l, \tilde {\rm f},
\tilde {\rm f}^*)\big|_{\delta\tilde l=0}}{}
 = \sum_{i=1}^N \big( \langle (\delta\gamma_n /\delta l)_+^*
f_i^*,\delta f_i\rangle + \langle (\delta\gamma_n /\delta l)_+ f_i,\delta f_i^*\rangle
\big)
\nonumber \\
\phantom{\delta\overline{\gamma}_n (\tilde l, \tilde {\rm f},
\tilde {\rm f}^*)\big|_{\delta\tilde l=0}}{}
 =\sum_{i=1}^N \big( \langle f_i^*, (\delta\gamma_n /\delta
l)_+\delta f_i\rangle  + \langle (\delta\gamma_n /\delta l)_+ f_i,\delta f_i^*\rangle
\big)
\nonumber \\
\phantom{\delta\overline{\gamma}_n (\tilde l, \tilde {\rm f},
\tilde {\rm f}^*)\big|_{\delta\tilde l=0}}{}
 =\sum_{i=1}^N \big( (\delta\gamma_n /\delta l,(\delta
f_i) \EE (\EE -1)^{-1} f_i^*)+ (\delta\gamma_n /\delta l, f_i \EE
(\EE -1)^{-1} \delta f_i^*) \big)
\nonumber \\
\phantom{\delta\overline{\gamma}_n (\tilde l, \tilde {\rm f},
\tilde {\rm f}^*)\big|_{\delta\tilde l=0}}{}
 =\left ( \delta\gamma_n /\delta l,\delta \sum_{i=1}^N
 f_i \EE (\EE -1)^{-1} f_i^* \right ) =
(\delta\gamma_n /\delta l,\delta l  ), \label{ohen19}
\end{gather}
where $\gamma_n\in I (\mathcal{G}^*) $, $n\in \mathbb{N}$ and the
brackets $\langle \cdot ,\cdot \rangle $ denote a scalar product on $W$.

As a result of expression \eqref{ohen19} one obtains the
relationship:
%%%18
\begin{gather}\label{ohen19a}
\left . \delta l \right |_{\delta \tilde l=0} =\sum_{i=1}^N \delta
\big(f_i \EE (\EE -1)^{-1} f_i^*\big) .
\end{gather}
From~\eqref{ohen19a} it follows directly that
\[
l=\mathcal{K}(\tilde l)+\sum_{i=1}^N f_i \EE (\EE -1)^{-1} f_i^* ,
\]
where $\mathcal{K}$ is an arbitrary Frechet-smooth operator on
$\mathcal{G}^*$. If $\mathcal{K}(\tilde l)=\tilde l$ for any
$\tilde l\in \mathcal{G}^*$ then
%%%19
\begin{gather}\label{ohen20}
l=\tilde l+\sum_{i=1}^N f_i \EE (\EE -1)^{-1} f_i^* .
\end{gather}
Thus, B\"acklund transformation \eqref{ohen17} can be written as
%%%20
\begin{gather}\label{ohen21}
(\tilde l, \tilde {\rm f}, \tilde {\rm f}^*)^\top
\stackrel{B}{\mapsto} ( l=\tilde l +\sum_{i=1}^N  f_i \EE (\EE
-1)^{-1} f_i^*, {\rm f}, {\rm f}^*) ^\top.
\end{gather}
The existence of B\"acklund transformation \eqref{ohen21} enables
the following theorem to be proved.
\begin{theorem}
Under the B\"acklund transformation \eqref{ohen21} the dynamical
system~\eqref{ohen18} on $\mathcal{G}^*\oplus W^{2N}$ is
equivalent to the system of evolution equations:
\begin{gather*}
d\tilde l/dt_n = [(\nabla\overline{\gamma}_n (\tilde l))_+,\tilde
l]-[\nabla \overline{\gamma}_n (\tilde l),\tilde l]_{>0} , \\
d\tilde {\rm f}/dt_n=\delta \overline{\gamma}_n /\delta\tilde {\rm
f}^*, \qquad d\tilde {\rm f}^*/dt_n=-\delta \overline{\gamma}_n
/\delta\tilde {\rm f},
\end{gather*}
where $\overline{\gamma}_n:=\left . \gamma_n \right |_{l=l(\tilde
l, {\rm f}, {\rm f}^*)} \in \mathcal{D} (\mathcal{G}^*\oplus
W^{2N})$ and $\gamma_n \in I(\mathcal{G}^*) $ is a Casimir
functional at the point $l\in \mathcal{G}^*$ for every $n\in
\mathbb{N}$.
\end{theorem}
The Frechet derivative $B^{\prime}: \TT(\mathcal{G}^*\oplus W^{2N})\to
\TT(\mathcal{G}^*\oplus W^{2N})$ of the B\"acklund
transformation~\eqref{ohen21} and the corresponding conjugate
operator $B^{\prime*}: \TT ^*(\mathcal{G}^*\oplus W^{2N})\to \TT
^*(\mathcal{G}^*\oplus W^{2N})$ take the following forms:
\begin{gather*}
\left ( \begin{array}{c} h \\ \alpha \\ \beta
\end{array} \right )
\stackrel{B^{\prime} }{\mapsto} \left ( \begin{array}{c}
h+\sum_{i=1}^N( \alpha_i \EE (\EE -1)^{-1} f_i^* +f_i \EE (\EE
-1)^{-1} \beta_i ) \\
\alpha \\ \beta
\end{array} \right ) ,
\\
\left ( \begin{array}{c} r \\ \chi \\ \rho
\end{array} \right )
\stackrel{B^{\prime*} }{\mapsto} \left ( \begin{array}{c} r \\
\chi+(r_+^* {\rm f}^*)
\\ \rho+(r_+ {\rm f})
\end{array} \right ) ,
\end{gather*}
where $(h,\alpha,\beta)^\top\in \TT_{(l,{\rm f},{\rm
f}^*)}(\mathcal{G}^*\oplus W^{2N})$ and $(r,\chi,\rho)^\top\in
\TT^*_{(l,{\rm f},{\rm f}^*)}(\mathcal{G}^*\oplus W^{2N})$ at the
point $(l,{\rm f},{\rm f}^*)^\top \in \mathcal{G}^*\oplus W^{2N}$,
$\alpha=(\alpha_1, \ldots, \alpha_N)^\top$, $\beta=(\beta_1,
\ldots, \beta_N)^\top$, $\chi=(\chi_1, \ldots, \chi_N)^\top$,
$\rho=(\rho_1, \ldots, \rho_N)^\top$.

By means of calculations via the
formula~(see~\cite{hentosh_Fokas,hentosh_Myk}):
\begin{gather}\label{ohen21a}
\Theta = B^{\prime} \tilde \Theta B^{\prime*},
\end{gather}
one f\/inds the B\"acklund transformed Poisson structure $\Theta $
on $\mathcal{G}^*\oplus W^{2N}$:
\begin{gather}\label{ohen22}
\nabla\overline\gamma (l, {\rm f}, {\rm f}^*)\stackrel{\Theta
}{\mapsto} \left ( \begin{array}{c} \left [ l,
(\delta\overline\gamma/\delta l)_+ \right ]-
\left [ l ,  \delta\overline\gamma/\delta l \right ]_{>0} +{}\\
{}+\sum\limits_{i=1}^N(f_i \EE (\EE -1)^{-1} (\delta\overline\gamma/\delta
f_i)- (\delta\overline\gamma/\delta f_i^*)\EE (\EE -1)^{-1} f_i^*
)
\vspace{1mm}\\
-\delta\overline\gamma/\delta {\rm f}^* -((\delta\overline\gamma/\delta l)_+{\rm f})
\vspace{1mm}\\
\delta\overline\gamma/\delta {\rm
f}+((\delta\overline\gamma/\delta l)^*_+{\rm f}^*)
\end{array} \right ),
\end{gather}
where $\overline\gamma\in \mathcal{D}(\mathcal{G}^*\oplus W^{2N})$
is an arbitrary Frechet-smooth functional. Thereby, one can
formulate the following theorem.
\begin{theorem}
The hierarchy of dynamical systems \eqref{ohen9}, \eqref{ohen13}
and \eqref{ohen14} is Hamiltonian with respect to the Poisson
structure $\Theta$ in the form~\eqref{ohen22} and the Casimir
functionals $\gamma_n\in I (\mathcal{G}^*) $, $n\in \mathbb{N}$,
as Hamiltonian functions.
\end{theorem}
Based on the expression~\eqref{ohen18} one can construct a new
hierarchy of Hamiltonian evolution equations generated by
involutive with respect to the Lie--Poisson bracket~\eqref{ohen3}
Casimir invariants $\gamma_n\in I(\mathcal{G}^*)$, $n\in
\mathbb{N}$, in the form~\eqref{ohen6} on the extended phase space
$\mathcal{G}^*\oplus W^{2N}$. On the coadjoint orbits of the Lie
algebra~$\mathcal{G}$ they give rise to the Lax representations
for some $(1+1)$-di\-men\-si\-onal dif\/ferential-dif\/ference
systems~\cite{hentosh16,shentosh1,hentosh_Bl,bhentosh6,shentosh_Kup,shentosh12b,shentosh18}.

\section[The additional symmetry hierarchies associated with the Lax integrable
(1+1)-dimensional differential-difference systems]{The additional symmetry hierarchies associated with the Lax\\ integrable
$\boldsymbol{(1+1)}$-dimensional dif\/ferential-dif\/ference systems}\label{section4}

The hierarchy of coupled evolution equations \eqref{ohen9},
\eqref{ohen13} and \eqref{ohen14} possesses another natural set of
invariants including all higher powers of the eigenvalues
$\lambda_k$, $k=\overline{1,N}$. They can be considered as
Frechet-smooth functionals on the extended phase space
$\mathcal{G}^*\oplus W^{2N}$, owing to the evident representation :
\begin{gather}\label{ohen40}
\lambda_k^s=\langle f_k^*,l^s f_k \rangle  , \qquad s\in \mathbb{N},
\end{gather}
holding for every $k=\overline{1,N}$ under the normalizing
constraint
\begin{gather*}%\label{ohen41a}
\langle f_k^*,f_k\rangle =1.
\end{gather*}
The Frechet-smooth functionals $\mu_i\in
\mathcal{D}(\mathcal{G}^*\oplus W^{2N})$, $i=\overline{1,N}$:
\[
\mu_i:=\langle f_i^*, f_i \rangle,
\]
are invariant with respect to the dynamical systems~\eqref{ohen9}, \eqref{ohen13}
and \eqref{ohen14}.

In the case of B\"acklund transformation~\eqref{ohen20}, where
\begin{gather}\label{ohen41}
l:= l_{>0}+\sum_{i=1}^N f_i\EE (\EE-1)^{-1} f_i^* ,
\end{gather}
formula \eqref{ohen40} gives rise to the following variation of
the functionals $\lambda_k^s\in \mathcal{D} (\mathcal{G}^*\oplus
W^{2N})$ for every $k=\overline{1,N}$ and $s\in \mathbb{N}$:
\begin{gather}
\delta \lambda_k^s= \langle \delta f_k^*,l^s f_k\rangle + \langle f_k^*,\delta (l^s
f_k)\rangle  +\langle f_k^*,l^s (\delta f_k)\rangle
\nonumber \\
\phantom{\delta \lambda_k^s}{} = (\delta (l^s)_{>0}, M_k^s) +\sum_{i=1}^N \langle \delta
f_i,(-M_k^s+\delta_k^i l^s)^*f_i^*\rangle
 +\sum_{i=1}^N \langle \delta f_i^*,(-M_k^s +\delta_k^i l^s)
f_i\rangle, \label{ohen42}
\end{gather}
where $\delta_k^i$ is the Kronecker symbol and the operators
$M_k^s$ are determined as
\[
M_k^s:= \sum_{p=0}^{s-1} ((l^pf_k)(\EE -1)^{-1}
((l^*)^{s-1-p}f_k^*)).
\]
It should be noted that the $s$th power of the
operator~\eqref{ohen41} takes the form
\[
l^s= (l^s)_{>0}+\sum_{i=1}^N \sum_{p=0}^{s-1} \big((l^pf_i)\EE (\EE
-1)^{-1} \big((l^*)^{s-1-p}f_i^*\big)\big).
\]
By means of the representation~\eqref{ohen42} one obtains the
exact forms of gradients for the functionals $\lambda_k^s\in
\mathcal{D} (\mathcal{G}^*\oplus W^{2N})$, $k=\overline{1,N}$:
\begin{gather}\label{ohen43}
\nabla \lambda_k^s (l_{>0},{\rm f}, {\rm f}^*) = (M_k^s,\,
(-M_k^s+\delta_k^i l^s)^*f_i^*,  (-M_k^s +\delta_k^i l^s) f_i :\
i=\overline{1,N} )^\top .
\end{gather}
The tensor product $\tilde \Theta$ of the Poisson structures
\eqref{ohen15} and \eqref{ohen16} together with the
relationships~\eqref{ohen43} generates a new hierarchy of coupled
evolution equations on $\mathcal{G}^*\oplus W^{2N}$:
%%%28,29,30
\begin{gather}
d l_{>0}/d\tau _{s,k}=-[M_k^s, l_{>0}]_{>0}  , \label{ohen44} \\
df_i/d\tau _{s,k} = (-M_k^s+\delta_k^i l^s) f_i ,  \label{ohen45} \\
df^*_i/d\tau _{s,k} = (M_k^s-\delta_k^i l^s)^*f_i^*,\label{ohen46}
\end{gather}
where $i=\overline{1,N}$ and $\tau_{s,k}\in \mathbb{R}$, $s\in
\mathbb{N}$, $k=\overline{1,N}$, are evolution parameters. Owing
to B\"acklund transformation \eqref{ohen21}, equation
\eqref{ohen44} can be rewritten as the following equivalent
commutator relationship:
\begin{gather}\label{ohen47}
d l/d\tau _{s,k}=-[M_k^s, l ] = -s\lambda_k^{s-1} [M_k^1, l]
=s\lambda_k^{s-1} dl/d\tau_{1,k} .
\end{gather}
Since the functionals $\mu_i\in \mathcal{D}(\mathcal{G}^*\oplus
W^{2N})$, $i=\overline{1,N}$, are invariant with respect to the
dynamical systems \eqref{ohen47}, \eqref{ohen45} and
\eqref{ohen46} one can formulate the following theorem.
\begin{theorem}
For every $k=\overline{1,N}$ and all $s\in \mathbb{N}$ the
Hamiltonian representations of the dynamical
systems~\eqref{ohen47}, \eqref{ohen45} and \eqref{ohen46} on their
invariant subspace $M_k\subset \mathcal{G}^*\oplus W^{2N}$:
\[
M_k:=\{ (l,{\mbox f},{\mbox f}^*)^\top \in \mathcal{G}^*\oplus
W^{2N}: \  \mu_k=1 \},
\]
are given by the Poisson structure $\Theta $ in the form
\eqref{ohen22} and the functionals $\lambda_k^s\in
\mathcal{D}(\mathcal{G}^*\oplus W^{2N})$ as Hamiltonian functions taken both to be reduced on~$M_k$.
\end{theorem}

On the subspace of the operators $l\in \mathcal{G}^*$ in the
forms~\eqref{ohen41} one has the following representation for the
f\/lows $d/dt_n$, $n\in \mathbb{N}$:
\begin{gather}\label{ohen48a}
d/dt_n=n\sum_{k=1}^N \lambda_k^{n-1}d/d\tau_{1,k}.
\end{gather}

\begin{theorem}
The dynamical systems \eqref{ohen47}, \eqref{ohen45} and
\eqref{ohen46} describe flows on $\mathcal{G}^*\oplus W^{2N}$
commuting both with each other and the hierarchy of Lax-type
dynamical systems \eqref{ohen9}, \eqref{ohen13} and
\eqref{ohen14}.
\end{theorem}

\begin{proof}
With taking into account the representation~\eqref{ohen48a} it is
suf\/f\/icient to show that
%%%
\[
[d/d \tau_{1,k}, d/d\tau_{1,q}]=0 ,
\]
where $k,q=\overline{1,N}$, $k\not = q$ and $n\in \mathbb{N}$.
This equality follows from the identity
\[
d M_k^1/d\tau_{1,q}-d M_q^1/d\tau_{1,k}=[M_k^1,M_q^1],
\]
holding because of the relationship
\begin{gather*}
M_k^1 M_q^1=(M_k^1f_q) (\EE-1)^{-1}f_q^*+f_k
(\EE-1)^{-1}((M_q^1)^*f_k^*) \\
\phantom{M_k^1 M_q^1}{} =-df_q/d\tau_{1,k} (\EE-1)^{-1}f_q^*+f_k
(\EE-1)^{-1}(df_k^*/d\tau_{1,q}) . \tag*{\qed}
\end{gather*}
\renewcommand{\qed}{}
\end{proof}

Thus, for every $k=\overline{1,N}$ and all $s\in\mathbb{N}$
dynamical systems \eqref{ohen47}, \eqref{ohen45} and
\eqref{ohen46} on $\mathcal{G}^*\oplus W^{2N}$ form a hierarchy of
additional homogeneous, or so called squared eigenfunction,
symmetries~\cite{hentosh_ANP,hentosh_Carillo1,hentosh_Carillo2,hentosh_NP,hentosh_Zeng1,hentosh_Zeng2}
for Lax-type f\/lows \eqref{ohen9}, \eqref{ohen13} and
\eqref{ohen14}.

Earlier the squared eigenfunction symmetry hierarchies associated
with the Lie algebras of integral-dif\/ferential and
super-integral-dif\/ferential operators in commutator forms as well
as their relations to some $(2+1)$-, $(2|1+1)$- and
$(2|2+1)$-dimensional Lax integrable nonlinear dynamical systems
on functional manifolds and supermanifolds were investigated
in~\cite{hentosh_ANP,hentosh_Carillo1,hentosh_Carillo2,bhentosh8,bhentosh27,shentosh9,hentosh_NP}.
In paper~\cite{hentosh_Zeng1} the commutator-type squared
eigenfunction symmetries were considered for the shift operator
Lie algebra.

In the case when $N\ge 2$ one can obtain a new class of nontrivial
independent f\/lows $d/dT_{n,K} :=d/dt_n +
\sum_{k=1}^{K}d/d\tau_{n,k}$, $K=\overline{1,[N/2]}$, $n\in
\mathbb{N}$, on $\mathcal{G}^*\oplus W^{2N}$ in the Lax-type forms
by use of the considered above invariants of the shift operator
Lie algebra $\mathcal{G}$. These f\/lows are Hamiltonian on their
invariant subspaces $\bigcap_{k=1}^K M_k\subset
\mathcal{G}^*\oplus W^{2N}$ because the following relationship
\[
\{\mu_i,\mu_q \}_{ \tilde J }=0,
\]
where $\mu_i\in \mathcal{D}(\mathcal{G}^*\oplus W^{2N})$ and
$\{\cdot,\cdot \}_{ \tilde J }$ is a Poisson bracket on $W^{2N}$, related
to the implectic operator~\eqref{ohen16}, holds for all
$i,q=\overline{1,N}$.

Acting on the eigenfunctions $(f_i,f^*_i)\in W^{2N}$,
$i=\overline{1,N}$, the f\/lows $d/dT_{n,K}$,
$K=\overline{1,[N/2]}$, $n\in \mathbb{N}$, generate some Lax
integrable $((1+K)+1)$-dimensional dif\/ferential-dif\/ference
dynamical systems. For example, in the case of the element
\[
l:=\EE+f_1 \EE (\EE -1)^{-1} f_1^*+f_2 \EE (\EE -1)^{-1} f_2^*\in
\mathcal{G}^*
\]
with $(f_1,f_2,f^*_1,f^*_2) \in W^4$ the f\/lows
$d/d\tau:=d/d\tau_{1,1}$ and $d/dT: =d/dT_{2,1}:
=d/dt_2+d/d\tau_{2,1}$ on $\mathcal{G}^*\oplus W^4$ give rise to
such dynamical systems as
%%%33
\begin{gather}
f_{1,\tau}=(\EE f_1)+f^2_1f_1^*+f_1f_2f_2^*+\bar u f_2 ,  \nonumber \\
f^*_{1,\tau}=-((\EE^{-1}f_1^*)+f_1(f_1^*)^2+f_1^*f_2f_2^*-(\EE u) f^*_2) ,\label{ohen49}  \\
f_{2,\tau}=-uf_1 , \qquad f^*_{2,\tau}=-(\EE\bar u)f^*_1 ,
\nonumber
\end{gather}
and
\begin{gather}
f_{1,T}=f_{1,\tau\tau}+(\EE^2f_1)+w_1(\EE f_1) +w_0f_1+2(f_1(\EE^{-1}f_1^*)+u\bar u)f_1  , \nonumber\\
f^*_{1,T}=-f^*_{1,\tau\tau}-(\EE^{-2}f^*_1)-(\EE w_1f^*_1)
-w_0f^*_1-2(f_1(\EE^{-1}f_1^*)+u\bar u)f^*_1  , \nonumber \\
f_{2,T}=(\EE^2f_2)+w_1(\EE f_2) +w_0f_2 -u f_{1,\tau}+ u_{\tau}f_1,
 \label{ohen50}\\
f^*_{2,T}=-(\EE^{-2}f^*_2)-(\EE w_1f^*_2) -w_0f^*_2
+\bar u f^*_{1,\tau} -  \bar u_{\tau}f^*_1 , \nonumber \\
(\EE-1)u=f_1^*f_2 , \qquad (\EE-1)\bar u =f_1f_2^* ,  \nonumber
\end{gather}
where one puts $(\nabla\gamma_2 (l))_+:=\EE^2+ w_1\EE +w_0$ and
$w_0,w_1, u, \bar u \in \mathcal{S} (\mathbb{Z};\mathbb{C})$ are
some functions, depending parametrically on variables $\tau, T\in
\mathbb{R}$.

Together systems \eqref{ohen49} and \eqref{ohen50} represent some
$(2+1)$-dimensional nonlinear dynamical system with an
inf\/inite sequence of conservation laws in the form~\eqref{ohen6}.
Its Lax-type linearization is given by spectral
problem~\eqref{ohen9} and the following evolution equations:
%%%35,36
\begin{gather}
f_{\tau}=-M_1^1f, \label{ohen51} \\
f_T=((\nabla\gamma_2 (l))_+-M_1^2)f, \label{ohen52}
\end{gather}
for an arbitrary eigenfunction $f\in W$. From the compatibility
condition of the relationships~\eqref{ohen51} and~\eqref{ohen52},
being equivalent to the commutability of f\/lows $d/d\tau$ and
$d/dT$ on the subspace of the operators $l\in \mathcal{G}^*$ in
the forms~\eqref{ohen41}, one has the equality
\[
d (\nabla \gamma_2(l))_+/d \tau_{1,k}=[(\nabla \gamma_2(l))_+,
M_1^1]_+ ,
\]
which leads to the additional nonlinear constraints
%%%37
\begin{gather*}
w_{0,\tau}=(\EE^2 f_1)f_1^*-f_1(\EE^{-2}f_1^*)+w_1 (\EE f_1)f_1^*-f_1 (\EE^{-1}w_1f_1^*) , \nonumber \\
w_{1,\tau}=(\EE^2 f_1)(\EE f_1^*)-f_1(\EE^{-1}f_1^*),%\label{ohen53}
\end{gather*}
for the dynamical system~\eqref{ohen49}, \eqref{ohen50}.

The results obtained in this section can be applied to
constructing a wide class of integrable $(2+1)$-dimensional
dif\/ferential-dif\/ference systems with triple Lax-type
linearizations.

\section[The Hamiltonian structure of the Lax integrable
(2+1)-dimensional differential-difference system on the
extended phase space]{The Hamiltonian structure of the Lax integrable\\
$\boldsymbol{(2+1)}$-dimensional dif\/ferential-dif\/ference system\\ on the
extended phase space}\label{section5}

One will assume that the Lie algebra $\mathcal{G}$ depends on the
parameter $y\in \mathbb{S}^1\simeq \mathbb{R}/2\pi\mathbb{Z}$,
which generates the loop Lie algebra $\mathcal{\hat G}:=
C^\infty(\mathbb{S}^1;\mathcal{G})$ with the $ad$-invariant
nondegenerate symmetric bilinear form:
\[
(A,B):= \int_0^{2\pi} dy\,  \mbox{Tr}\,(AB), \qquad A,B\in
\mathcal{\hat G}.
\]
The Lie algebra $\mathcal{\hat G}$ can be extended via the central
extension
procedure~\cite{shentosh2,shentosh3,shentosh5,bhentosh6,bhentosh10,hentosh4}
to the Lie algebra $\mathcal{\hat G}_c:= \mathcal{\hat G}\oplus
\mathbb{C}$ with the commutator
\[
[(A,\alpha),(B,\beta)]=([A,B],\omega_2(A,B)),\qquad A,B\in
\mathcal{\hat G}, \qquad \alpha,\beta\in \mathbb{C},
\]
where $\omega_2(\cdot,\cdot)$ is a standard Maurer--Cartan 2-cocycle on
$\mathcal{\hat G}$ such that
\[
\omega_2(A,B):=(A,\partial B/\partial y), \qquad A,B\in
\mathcal{\hat G},
\]
and the scalar product takes the form
%%%38
\begin{gather}\label{ohen23}
((A,\alpha),(B,\beta)):=(A,B)+\alpha\beta.
\end{gather}
The $\mathcal{R}$-deformed
commutator~\cite{shentosh2,shentosh3,shentosh5,bhentosh6,bhentosh10,hentosh4}
on $\mathcal{\hat G}_c$ such that
\[
[(A,\alpha),(B,\beta)]_{\mathcal R}=([A,B]_{\mathcal
R},\omega_{2,\mathcal R}(A,B)), \qquad A,B\in \mathcal{\hat G},
\]
where
\[
\omega_{2,\mathcal R}(A,B)=\omega_2(\mathcal
RA,B)+\omega_2(A,\mathcal RB),
\]
leads to the Lie--Poisson bracket
%%%39
\begin{gather}\label{ohen24}
\left\{ \gamma,\mu \right\}_{\mathcal R}(l)= \left (
l,[\nabla\gamma(l),\nabla\mu(l)]_{\mathcal{R} }  \right )
+c\omega_{2,\mathcal R}(\nabla\gamma(l),\nabla\mu(l)),
\end{gather}
where $\gamma,\mu\in \mathcal{D}(\mathcal{\hat G}^*_c)$ are some
Frechet-smooth functionals, $l\in\mathcal{\hat G}^*$ and
$c\in\mathbb{C}$, on the dual space $\mathcal{\hat G}^*_c\simeq
\mathcal{\hat G}_c$ to $\mathcal{\hat G}_c$ with respect to the
scalar product~\eqref{ohen23}.

The corresponding Casimir functionals $\gamma_n\in I(\mathcal{\hat
G}^*_c)$, $n\in \mathbb{N}$, obey the relationship:
\[
[l-c\partial/\partial y , \nabla\gamma (l)]=0,
\]
which can be solved by use of the following
expansions~\cite{hentosh4}
%%%40
\begin{gather}\label{ohen25}
\nabla\gamma_n (l):=\sum_{n-j\in \mathbb{Z}_+}u_j \EE ^j.
\end{gather}
The Lie--Poisson bracket~\eqref{ohen24} together with the Casimir
functionals $\gamma_n\in I(\mathcal{\hat G}^*_c)$, $n\in
\mathbb{N}$, generates the Hamiltonian f\/lows
%%%41
\begin{gather}\label{ohen26}
d l/dt_n:= [\mathcal{R}\nabla\gamma_n (l),\   l- c\partial
/\partial y]= [(\nabla\gamma_n (l))_+,\  l- c\partial /\partial
y].
\end{gather}
Every Hamiltonian f\/low in~\eqref{ohen26} can be considered as a
compatibility condition for the spectral problem
%%%42
\begin{gather}\label{ohen27}
((l - c\partial /\partial y ) f)= \lambda f,
\end{gather}
where $f\in \bar W:=L_2 (\mathbb{Z}\times
\mathbb{S}^1;\mathbb{C})$, $\lambda\in\mathbb{C}$ is a spectral
parameter, and the following evolution equation:
%%%43
\begin{gather}\label{ohen28}
df/dt_n= ((\nabla\gamma_n (l))_+f).
\end{gather}
The corresponding evolution for the adjoint eigenfunction $f^*\in
\bar W^*\simeq \bar W$ takes the form:
%%%44
\begin{gather}\label{ohen29}
df^*/dt_n=-((\nabla\gamma_n (l))^*_+ f^*).
\end{gather}
One will investigate the existence problem of a Hamiltonian
representation for equation~\eqref{ohen26} coupled with $N\in
\mathbb{N}$ copies of~\eqref{ohen28}:
%%%45
\begin{gather}\label{ohen30}
df_i/dt_n= ((\nabla\gamma_n (l))_+f_i),
\end{gather}
for the corresponding eigenfunctions $f_i\in \bar W$,
$i=\overline{1,N}$, and the same number of copies
of~\eqref{ohen29}:
%%%46
\begin{gather}\label{ohen31}
df_i^*/dt_n=-((\nabla\gamma_n (l))^*_+ f_i^*),
\end{gather}
for the suitable adjoint eigenfunctions $f_i^*\in \bar W^*$, in
the case when the spectral relationship~\eqref{ohen27} admits $N
\in \mathbb{N}$ dif\/ferent eigenvalues $\lambda_i\in \mathbb{C}$,
$i=\overline{1,N}$.

On the spaces $\mathcal{\hat G}^*_c$ and $\bar W^{2N}$ there exist
canonical Poisson structures in the corresponding forms
\begin{gather}\label{ohen32}
\delta\overline\gamma/\delta\tilde
l\stackrel{\tilde\theta}{\mapsto} [\tilde l -c\partial /\partial
y,(\delta\overline\gamma/\delta\tilde l)_+] - [\tilde l-c\partial
/\partial y,\delta\overline\gamma/\delta\tilde l]_{>0} ,
\end{gather}
where $\tilde\theta: \TT^*(\mathcal{\hat G}^*_c)\to
\TT(\mathcal{\hat G}^*_c)$ is an implectic operator related to
\eqref{ohen24} at the point $\tilde l\in\mathcal{\hat G}^*$, and
%%%48
\begin{gather}\label{ohen33}
(\delta\overline\gamma/\delta\tilde {\rm f},\,
\delta\overline\gamma/\delta\tilde {\rm f}^*)^\top
\stackrel{\tilde J}{\mapsto} (-\delta\overline\gamma/\delta\tilde
{\rm f}^*,\, \delta\gamma/\delta\tilde {\rm f})^\top ,
\end{gather}
where $\tilde J:\TT^*(\bar W^{2N})\to \TT(\bar W^{2N})$ at the
point $(\tilde {\rm f}, \tilde {\rm f}^*)\in \bar W^{2N}$, for any
Frechet-smooth functional $\overline\gamma\in
\mathcal{D}(\mathcal{\hat G}^*_c\oplus \bar W^{2N})$.

The tensor product $\tilde\Theta :=\tilde\theta\otimes \tilde J$
of \eqref{ohen32} and \eqref{ohen33} can be considered as a
Poisson structure on the extended phase space $\mathcal{\hat
G}^*_c\oplus \bar W^{2N}$.

Applying the procedure described in Section~\ref{section3} to the coupled
dynamical system~\eqref{ohen26}, \eqref{ohen30} and \eqref{ohen31}
one obtains a B\"acklund transformation on the extended phase
space $\mathcal{\hat G}^*_c\oplus \bar W^{2N}$ in the
form~\eqref{ohen21}. Thus, the following theorem can be
formulated.
\begin{theorem}
Under the B\"acklund transformation \eqref{ohen21} the dynamical
system~\eqref{ohen18} on $\mathcal{\hat G}^*_c\oplus \bar W^{2N}$
is equivalent to the system of evolution equations:
\begin{gather*}
d\tilde l/dt_n = [(\nabla\overline{\gamma}_n (\tilde l))_+, \tilde
l]- [\nabla \overline{\gamma}_n (\tilde l),\tilde l]_{>0} +c
(\partial /\partial y
(\nabla\overline{\gamma}_n (\tilde l))_0), \\
d\tilde {\rm f}/dt_n=\delta \overline{\gamma}_n /\delta\tilde {\rm
f}^*, \qquad  d\tilde {\rm f}^*/dt_n=-\delta \overline{\gamma}_n
/\delta\tilde {\rm f} ,
\end{gather*}
where $\overline{\gamma}_n:=$ $\left . \gamma_n \right
|_{l=l(\tilde l, {\rm f}, {\rm f}^*)} $ $\in \mathcal{D}
(\mathcal{\hat G}^*_c\oplus \bar W^{2N}) $ and $\gamma_n \in I
(\mathcal{\hat G}^*_c) $ is a Casimir functional at the point
$l\in \mathcal{G}^*$ for every $n\in \mathbb{N}$.
\end{theorem}
By means of calculations via the formula~\eqref{ohen21a} one can
f\/ind the following form of the B\"acklund transformed Poisson
structure $\Theta $ on $\mathcal{\hat G}^*_c\oplus \bar W^{2N}$
%%%49
\begin{gather}\label{ohen34}
\nabla\overline\gamma (l, {\rm f}, {\rm f}^*)\stackrel{\Theta
}{\mapsto} \left ( \begin{array}{c} \left [ l-c\partial /\partial
y, (\delta\overline\gamma/\delta l)_+  \right ]-
\left [ l-c\partial /\partial y,  \delta\overline\gamma/\delta l \right ]_{>0} +{}\\
{}+\sum\limits_{i=1}^N(f_i \EE (\EE -1)^{-1} (\delta\overline\gamma/\delta
f_i)- (\delta\overline\gamma/\delta f_i^*) \EE (\EE -1)^{-1} f_i^*
)
\vspace{1mm}\\
-\delta\overline\gamma/\delta {\rm f}^* -((\delta\overline\gamma/\delta l)_+{\rm f}) \vspace{1mm}\\
\delta\overline\gamma/\delta {\rm
f}+((\delta\overline\gamma/\delta l)^*_+{\rm f}^*)
\end{array} \right ) ,
\end{gather}
where $\overline\gamma\in \mathcal{D}(\mathcal{\hat G}^*_c\oplus
\bar W^{2N})$ is an arbitrary Frechet-smooth functional. Thus, the
following theorem holds.
\begin{theorem}
The hierarchy of dynamical systems \eqref{ohen26}, \eqref{ohen30}
and \eqref{ohen31} is Hamiltonian with respect to the Poisson
structure $\Theta$ in the form \eqref{ohen34} and the Casimir
functionals $\gamma_n\in I (\mathcal{\hat G}^*_c) $, $n\in
\mathbb{N}$, as Hamiltonian functions.
\end{theorem}

On the coadjoint orbits of the Lie algebra~$\mathcal{\hat
G}_c$ the Hamiltonian evolution equations~\eqref{ohen18}
gene\-ra\-ted
by Casimir invariants $\gamma_n\in I(\mathcal{\hat
G}^*_c)$, $n\in \mathbb{N}$, in the form~\eqref{ohen25}, being
involutive with respect to the Lie--Poisson bracket~\eqref{ohen24}, on the
extended
phase space $\mathcal{\hat
G}^*_c\oplus W^{2N}$ give rise to the Lax representations for some
$(2+1)$-di\-men\-si\-onal dif\/ferential-dif\/ference
systems~\cite{shentosh2,bhentosh10,shentosh10,hentosh4}. The
Lie-algebraic structure of the corresponding Hamiltonian
additional homogeneous symmetry hierarchies can be described by means of the
approaches developed in Section~\ref{section4} and in the
paper~\cite{bhentosh27}.

\section{Conclusion}\label{section6}

In this paper the method of solving the existence problem of
Hamiltonian representations for the coupled Lax-type hierarchies
on extended phase spaces which was proposed
in~\cite{bhentosh8,bhentosh27,shentosh9,hentosh12} have been
developed for Lax integrable $(1+1)$- and $(2+1)$-dimensional
dif\/ferential-dif\/ference systems of a lattice type associated with
the Lie algebra of shift operators and its central extension by
the Maurer--Cartan 2-cocycle correspondingly.

It is based on the invariance property of Casimir functionals under
some specially constructed B\"acklund transformation on a dual
space to the related operator Lie algebra whose a structure is
strongly depending on a Lie algebra splitting into a direct sum of
Lie subalgebras. Another
possibility~\cite{shentosh1,bhentosh10,hentosh_Oevel1,hentosh4} of
choosing a such splitting can give rise to a dif\/ferent B\"acklund
transformation.

For the coupled Lax-type hierarchy on the extended phase space of
the shift operator Lie al\-geb\-ra the Hamiltonian representations for
the additional homogeneous symmetry hierarchies have been obtained by using
the B\"acklund transformation mentioned above. It was shown that
these hierarchies generate a new class of $(2+1)$-dimensional
dif\/ferential-dif\/ference systems of a lattice type which possess
inf\/inite sequences of conservation laws and triple Lax-type
linearizations. The latter makes it possible to apply the
reduction procedure upon invariant solution
subspaces~\cite{hentosh_Bl,bhentosh10,shentosh12a,shentosh12b}.

The approaches described in the paper can be used to solve
analogous problems for nonlocal dif\/ferential-dif\/ference
systems~\cite{shentosh1,shentosh6,bhentosh10,hentosh4}.

For the considered classes of Lax integrable lattice and nonlocal
dif\/ferential-dif\/ference systems it is still an open problem to
develop the appropriate Darboux--B\"acklund transformation
method~\cite{bhentosh14,bhentosh24} which has been ef\/fective for solving Lax integrable nonlinear dynamical
systems on functional manifolds.

\pdfbookmark[1]{References}{ref}
\LastPageEnding

\end{document}